\documentclass[conference,anonymous,review]{IEEEtran}
\usepackage{graphicx} % Required for inserting images
\usepackage{ulem}
\usepackage{xcolor}
\usepackage[flushleft]{threeparttable}
\usepackage{subcaption}
\usepackage[linesnumbered, ruled, vlined]{algorithm2e}
\usepackage{amssymb}
\usepackage{amsmath}

\usepackage{xcolor}
\usepackage{microtype}
\usepackage{cuted}
\usepackage{verbatim}
\usepackage{comment}

\usepackage{booktabs}
\usepackage{array}
\usepackage{hyperref}
\usepackage{xcolor}
\usepackage{multirow}
\usepackage{caption}
\usepackage{enumitem}
\usepackage{nicefrac}
\usepackage{mdframed}
\usepackage{tcolorbox}
\usepackage{longtable}
\usepackage{float}
\tcbuselibrary{skins,breakable}
\usepackage{makecell}

\usepackage{booktabs}       % for \toprule, \bottomrule
\usepackage{threeparttable} % for threeparttable + tablenotes
\usepackage{xcolor}
\def\BibTeX{{\rm B\kern-.05em{\sc i\kern-.025em b}\kern-.08em
    T\kern-.1667em\lower.7ex\hbox{E}\kern-.125emX}}
    
\newcommand{\dmodel}{d_{\mathrm{model}}}
\newcommand{\dffnmoe}{d_{\mathrm{ffn}}^{\mathrm{MoE}}}
\newcommand{\dffnden}{d_{\mathrm{ffn}}^{\mathrm{d}}}
\newcommand{\betap}{\beta_p}

\definecolor{phase1col}{RGB}{30, 100, 200}
\definecolor{phase2col}{RGB}{180, 40, 40}
\definecolor{phase3col}{RGB}{30, 150, 80}
\definecolor{commentcol}{RGB}{120, 120, 120}

\SetCommentSty{textnormal}
\newcommand{\mycomment}[1]{%
  \textcolor{commentcol}{\textit{\scriptsize // #1}}}
\SetKwComment{Comment}{\mycomment{}{}}{}
\SetArgSty{textnormal}
\SetAlFnt{\small}
\SetAlCapFnt{\small}
\SetAlCapNameFnt{\small}

\SetKwBlock{PhaseOne}%
  {\textcolor{phase1col}{\textbf{Phase 1:}} Intra-node Extraction}{}
\SetKwBlock{PhaseTwo}%
  {\textcolor{phase2col}{\textbf{Phase 2:}} Inter-node Exchange}{}
\SetKwBlock{PhaseThree}%
  {\textcolor{phase3col}{\textbf{Phase 3:}} Intra-node Redistribution}{}

\newcommand{\sh}[1]{\ensuremath{(#1)}}

\begin{document}
\title{Piper: Efficient Large-Scale MoE Training via Resource Modeling and Pipelined Hybrid Parallelism}
%\author{Sajal Dash}

\author{
    \IEEEauthorblockN{Sajal Dash}
    \IEEEauthorblockA{Oak Ridge National Laboratory\\
    Oak Ridge, USA\\
    Email: dashs@ornl.gov}
    \and
    \IEEEauthorblockN{Feiyi Wang}
    \IEEEauthorblockA{Oak Ridge National Laboratory\\
    Oak Ridge, USA\\
    Email: fwang2@ornl.gov}
}
\date{May 2026}

\maketitle
\begin{abstract}
Frontier models increasingly adopt Mixture-of-Experts (MoE) architectures to achieve large-model performance at reduced cost. However, training MoE models on HPC platforms is hindered by large memory footprints, frequent large-scale communication across heterogeneous networks, and severe workload imbalance. To characterize these challenges, we develop a mathematical model that quantifies memory, compute, and communication requirements for MoE configurations under various parallelization schemes, verified through micro-benchmarking, code instrumentation, and hardware profiling. Our analysis identifies performance bottlenecks: all-to-all latency at scale from expert parallelism, insufficient compute-communication overlap, low GPU utilization from imbalanced skinny GEMMs, and the absence of platform-aware hybrid parallelization strategies. To address these, we introduce Piper, a framework that leverages resource modeling to identify efficient training strategies for MoE models on target HPC platforms, applying pipeline parallelism with optimized schedules. Piper achieves 2–3.5× higher MFU than state-of-the-art frameworks such as X-MoE, and a novel all-to-all algorithm delivers 1.2X–9X bandwidth over vendor implementation. 
\end{abstract}

\section{Introduction}
Transformer based Large Language Models (LLMs) have demonstrated consistent performance gains with increasing model size~\cite{kaplan2020scaling, vaswani2017attention}. However, training these models at scale is resource-intensive: a mixed-precision training run requires approximately 20 bytes of memory per parameter, roughly 6 floating-point operations per parameter per token, and substantial inter-device communication~\cite{narayanan2021efficient, rajbhandari2020zero}. The Mixture-of-Experts (MoE) mechanism, in which only a fraction of the model parameters get sparsely activated for each token, has emerged as a promising approach to achieving dense-models' performance at a reduced computational cost with savings proportional to the sparsity factor~\cite{shazeer2017outrageously, fedus2022switch}. State-of-the-art models such as Mixtral~\cite{jiang2024mixtral}, DeepSeek~\cite{deepseekai2025deepseekr1}, Qwen~\cite{qwen_moe}, and Kimi~\cite{kimi2025} have adopted MoE to deliver superior performance at substantially reduced training cost.
Despite these advantages, MoE training presents its own unique challenges. Relative to a parameter-matched dense model, MoE training introduces computational load imbalance across devices, an elevated memory footprint from storing both model parameters and transient activation tensors, and high inter-device communication volume under expert parallelism — the most widely used MoE distribution method~\cite{lepikhin2021gshard, rajbhandari2022deepspeedmoe}. Emerging MoE architectures with fine-grained experts such as DeepSeek-MoE~\cite{dai2024deepseekmoe}, exacerbate these challenges further: they produce many tall-and-skinny GEMMs with poor hardware utilization, inflate activation memory, and require all-to-all collectives involving a large number of participating processes.

These difficulties compound when training on shared HPC platforms which are primarily designed for modeling and simulation workloads and feature non-uniform communication fabrics across GPU nodes. Prior work such as X-MoE~\cite{xmoe2024} has demonstrated that communication overhead becomes a dominant performance bottleneck in exactly this setting, yet no holistic framework exists for systematically characterizing, quantifying, and mitigating these inefficiencies across both model and platform architecture.

In this work, we develop mathematical models to quantify memory, compute, and communication requirements for diverse MoE architectures under different distributed training strategies and their combinations, parameterized by empirically measured platform characteristics — including memory capacity, GPU throughput, and network bandwidth. We validate these models through comprehensive experimental training runs using frameworks including DeepSpeed-TED~\cite{deepspeed_ted}, DeepSpeed-MoE~\cite{rajbhandari2022deepspeedmoe}, and X-MoE~\cite{xmoe2024} on the Frontier supercomputer~\cite{frontier2023}.

Using our resource modeling framework alongside experimental profiling, we identify prominent sources of missed performance: (i) all-to-all latency at scale caused by expert parallelism over non-uniform interconnects;  (ii) load imbalance from skewed expert assignment, resulting in low GPU utilization during early stages, and, in case of no external mechanisms, throughout the majority of training; and (iii) a lack of system-aware hybrid parallelization strategies. We then develop targeted solutions for each.
\subsection{Contributions}
\begin{enumerate}
\item \textbf{Analytical and Empirical Resource Modeling:} We develop a mathematical model for estimating memory, compute, and communication utilization during MoE training across a range of architectures and parallelization configurations, and empirically validate it through micro-benchmarking, code instrumentation, and hardware profiling.
\item \textbf{Piper: Pipeline Parallelism for Localizing Communication:} We introduce Piper, a framework that applies pipeline parallelism to intra-layer parallelization strategies in order to localize and overlap expensive collective communications, and uses the resource model to automatically identify efficient training configurations for target HPC platforms. Piper achieves 2–3.5X higher Model FLOP Utilization (MFU) compared to state-of-the-art MoE training framework, X-MoE~\cite{xmoe2024}.

\item \textbf{Topology-Aware All-to-All Algorithm:} We design a Dragonfly-topology-aware hierarchical all-to-all algorithm that exploits the dependency structure of asynchronous point-to-point communications, groups traffic over slower inter-node and inter-cabinet links, and saturates NICs uniformly, achieving 1.5X–4X the bandwidth of vendor-provided implementations.

\item \textbf{Expert Migration for Load Balancing:} We develop an expert migration technique in which GPUs hosting the same layer periodically exchange experts to re-balance load, incurring an amortized overhead of less than 5\% of total training time.

\item \textbf{Trillion-Scale MoE Training:} Using Piper and our resource modeling tool, we devise and validate training strategies for several state-of-the-art MoE models at 20–50\% MFU, and demonstrate training of trillion-parameter MoE models at $20\%$ MFU on the Frontier supercomputer. For reference, X-MoE reported training of a 545B parameter model at $5.23\%$ MFU.
\end{enumerate}

\section{Background and Related Work}
We survey prior work across five dimensions relevant to this paper: MoE model architectures, distributed training frameworks and hybrid parallelism, collective communication algorithms, load balancing techniques, and pipeline parallelism. We also review HPC platform topology characteristics that motivate our system-level design choices. Table~\ref{tab:moe-models} summarizes the architectural parameters of representative state-of-the-art MoE models.

\subsection{MoE Architectures}
The MoE mechanism replaces the feed-forward network (FFN) sublayer in a Transformer block~\cite{vaswani2017attention} with a collection of expert FFNs, each receiving only the tokens routed to it by a learned gating function~\cite{shazeer2017outrageously}. There are two primary architectural streams.
\textbf{Coarse-grained MoE.} Early large-scale MoE models such as GShard~\cite{lepikhin2021gshard}, Switch Transformer~\cite{fedus2022switch}, and the Mixtral family~\cite{jiang2024mixtral} employ a small number (typically 8–64) of large experts whose FFN dimension matches a comparably sized dense model, with top-1 or top-2 routing per token. Because individual experts frequently exceed single-GPU memory capacity, they require tensor parallelism or sharded data parallelism, substantially complicating the communication pattern.

\begin{table*}[ht]
\centering
\caption{State-of-the-Art Mixture-of-Experts (MoE) Model Configurations}
\label{tab:moe-models}
\resizebox{\linewidth}{!}{%
\begin{tabular}{@{}lrrccrrrll@{}}
\toprule
\textbf{Model} &
\textbf{Total Params} &
\textbf{Active Params} &
\textbf{Total Experts} &
\textbf{Active / Token} &
\textbf{Layers} &
\textbf{Hidden Size} &
\textbf{FFN Dim (per expert)} &
\textbf{Context} &
\textbf{Train Tokens} \\
\midrule
 
%\multicolumn{10}{@{}l}{\textit{DeepSeek}} \\[2pt]
DeepSeek-V2     & 236B  & 21B  & $162^{a}$  & 6R\,+\,2S & 60 & 5,120 & 1,536  & 128K  & 8.1T  \\
DeepSeek-V3     & 671B  & 37B  & $257^{b}$  & 8R\,+\,1S & 61 & 7,168 & 2,048  & 128K  & 14.8T \\
DeepSeek-V3.2   & 671B  & 37B  & $257^{b}$  & 8R\,+\,1S & 61 & 7,168 & 2,048  & 128K  & ---   \\
\midrule
 
%\multicolumn{10}{@{}l}{\textit{Mistral AI}} \\[2pt]
Mixtral $8{\times}7$B   & ${\sim}47$B$^{c}$ & ${\sim}13$B & 8 & Top-2 & 32 & 4,096 & 14,336 & 32K  & --- \\
Mixtral $8{\times}22$B  & 141B              & 39B         & 8 & Top-2 & 56 & 6,144 & 16,384 & 64K  & --- \\
\midrule
 
%\multicolumn{10}{@{}l}{\textit{Alibaba / Qwen}} \\[2pt]
Qwen3-30B-A3B   & 30B   & 3B   & $128^{d}$ & Top-8 & 48 & 2,048 & 768    & 128K  & ${\sim}36$T \\
Qwen3-235B-A22B & 235B  & 22B  & $128^{d}$ & Top-8 & 94 & 7,168 & 2,048  & 128K  & ${\sim}36$T \\
\midrule
 
%\multicolumn{10}{@{}l}{\textit{Meta}} \\[2pt]
Llama 4 Scout    & 109B & 17B & $17^{e}$   & 1R\,+\,1S & ${\sim}48$ & ${\sim}5{,}120$ & ${\sim}8{,}192$ & 10M & 40T \\
Llama 4 Maverick & 400B & 17B & $129^{e}$  & 1R\,+\,1S & ${\sim}48$ & ${\sim}5{,}120$ & ${\sim}8{,}192$ & 1M  & 40T \\
\midrule
 
%\multicolumn{10}{@{}l}{\textit{Snowflake}} \\[2pt]
Arctic           & 480B & 17B & $128^{f}$ & Top-2 & --- & --- & ${\sim}3{,}660$ & 128K & 3.5T \\
\midrule
 
%\multicolumn{10}{@{}l}{\textit{Moonshot AI}} \\[2pt]
Kimi K2          & ${\sim}1$T & 32B & 384 & Top-8 & 61 & 7,168 & 2,048 & 128K & 15.5T \\
 
\bottomrule
\end{tabular}%
}
\vspace{4pt}
\begin{minipage}{\linewidth}
\footnotesize
$^{a}$~160 routed (R) + 2 shared (S) experts.\quad
$^{b}$~256 routed + 1 shared expert.\quad
$^{c}$~Mixtral replicates only the FFN layers; attention weights are shared, giving ${\sim}47$B total rather than $8\times7\text{B}=56$B.\\
$^{d}$~No shared experts; uses global-batch load-balancing loss.\quad
$^{e}$~Maverick uses alternating dense and MoE layers; values marked ${\sim}$ are approximate (not fully disclosed by Meta).\\
$^{f}$~Arctic is a Dense-MoE hybrid: 10B dense transformer backbone + residual $128\times3.66$B MoE MLP.\quad
R\,=\,routed, S\,=\,shared.
\end{minipage}
\end{table*}

\textbf{Fine-grained MoE.} Pioneered by DeepSeek-MoE~\cite{dai2024deepseekmoe} and followed by Qwen3~\cite{qwen3}, Kimi K2~\cite{kimik2_2025}, and others, this approach decomposes each expert into many smaller experts; reducing FFN dimension by a factor $m$ while increasing expert count proportionally and selects a larger top-K ($K \in [6, 8, 16]$).
16]) per token. While individual experts now fit within a single GPU without tensor splitting, fine-grained designs generate tall-and-skinny GEMMs with poor hardware utilization, inflate activation memory by factor mm
m, and require all-to-all collectives spanning many GPUs.

\subsection{Distributed Training Frameworks and Hybrid Parallelism}
Training large MoE models requires combining data parallelism (DP)~\cite{rajbhandari2020zero}, tensor parallelism (TP)~\cite{narayanan2021megatron}, pipeline parallelism (PP)~\cite{narayanan2019pipedream, narayanan2021memory}, and expert parallelism (EP)~\cite{lepikhin2021gshard}, where different experts are placed on different devices and tokens are routed via all-to-all collectives.

\textbf{DeepSpeed-MoE}~\cite{rajbhandari2022deepspeedmoe} combines expert parallelism with tensor parallelism and ZeRO memory sharding~\cite{rajbhandari2020zero}, primarily targeting coarse-grained architectures. DeepSpeed-TED~\cite{deepspeed_ted} extends this by jointly optimizing across Tensor, Expert, and Data parallelism axes.

\textbf{X-MoE}~\cite{xmoe2024} targets fine-grained expert architectures, identifying activation memory and all-to-all scope as primary bottlenecks. It introduces zero-padding for load balancing, redundancy-based communication bypassing, and sequence-sharded parallelism, successfully outperforming DeepSpeed-MoE and Tutel at hundreds-of-billions scale. However, for 500B+ models, X-MoE achieves only $5\%$ MFU.

\textbf{Tutel}~\cite{tutel2022} provides efficient MoE dispatch and combine kernels with dynamic top-K routing and adaptive parallelism switching, but focuses on the dispatch kernel rather than end-to-end training strategy selection and does not cover attention-layer parallelization.

A general limitation across these frameworks is the absence of platform-aware hybrid parallelism planning that jointly accounts for memory, compute, and communication constraints, a gap Piper directly addresses (Section~\ref{sec:model}).

\subsection{Pipeline Parallelism}
Pipeline parallelism partitions model layers across devices using micro-batching~\cite{narayanan2019pipedream}. The 1F1B schedule~\cite{narayanan2021memory} reduces pipeline bubble fraction and peak activation memory over GPipe~\cite{huang2019gpipe}; interleaved variants~\cite{narayanan2021efficient} and ZB-H1/H2 schedules~\cite{qi2023zero} reduce bubble overhead further. These techniques target dense models where communication occurs between layers rather than within them. Piper extends pipeline parallelism to the intra-layer axis introduced by expert parallelism, enabling computation-communication overlap within MoE layers.

\subsection{Load Balancing}
Uneven token distribution across experts reduces effective GPU throughput. Common mitigations include auxiliary load-balancing losses~\cite{fedus2022switch, lepikhin2021gshard}, token dropping~\cite{fedus2022switch}, expert-choice routing~\cite{zhou2022mixture}, and the auxiliary-loss-free bias-adjustment strategy of DeepSeekV3~\cite{deepseekai2025deepseekr1}. All operate at the routing level and cannot correct device-level imbalance from oblivious expert placement. Our expert migration approach (Section~\ref{sec:migration}) complements these methods by physically redistributing experts based on observed load.

\subsection{Collective Communication Algorithms}
Under expert parallelism, each MoE layer incurs four all-to-all operations (two per forward pass and two per backward), making all-to-all a dominant latency contributor at scale~\cite{xmoe2024, rajbhandari2022deepspeedmoe}.

\textbf{Flat all-to-all.} NCCL~\cite{nccl} and RCCL~\cite{rccl} perform direct point-to-point transfers between all process pairs, which is bandwidth-optimal under a uniform topology but performs poorly on hierarchical networks where inter-node bandwidth is significantly lower than intra-node bandwidth.

\textbf{Hierarchical all-to-all.} Tutel~\cite{tutel2022}, FasterMoE~\cite{he2022fastermoe}, and HetuMoE~\cite{hetumoe2022} use a two-phase approach, intra-node aggregation followed by reduced inter-node transfers, substantially reducing cross-node messages. However, these algorithms treat the inter-node network as homogeneous.

\textbf{Dragonfly topology.} HPC systems such as Frontier~\cite{frontier2023} employ Dragonfly networks~\cite{kim2008dragonfly} with high-bandwidth intra-group links and sparser inter-group links. Topology-oblivious algorithms cause unnecessary contention on slower inter-group links. Our topology-aware all-to-all (Section~\ref{sec:alltoall}) explicitly models this three-level hierarchy (intra-node, intra-group, inter-group) and coordinates asynchronous point-to-point communication to eliminate idle cycles.

\subsection{Analytical Performance Modeling}
Roofline analysis~\cite{williams2009roofline} bounds achievable performance by arithmetic intensity and memory bandwidth, informing kernel optimization for attention and FFN layers~\cite{dao2022flashattention}. Korthikanti et al.~\cite{korthikanti2023reducing} develop analytical models for pipeline bubble and activation memory in dense Transformer training; PaLM~\cite{chowdhery2022palm} uses empirical roofline fitting to project hardware efficiency across configurations.
For MoE models, interactions among expert parallelism, routing, and load imbalance introduce variables that dense-model frameworks do not capture. No prior work provides a unified model spanning memory, compute, and communication jointly across different parallelization dimensions validated on a real HPC platform, a gap we address in Section~\ref{sec:model}.
\section{Piper: A Framework for Pipelining MoE Training via Resource Modeling}
\label{sec:model}

Piper is a framework for efficient MoE model training on HPC platforms. Its design rests on two observations. First, existing frameworks such as  DeepSpeed-MoE, DeepSpeed-TED, and X-MoE, distribute all model components across large, groups of GPUs, forcing expensive collective communications (4 all-reduce for tensor parallelism, 2 all-gather for sharded data parallelism, 4 all-to-all for expert parallelism) to span many ranks simultaneously. Second, pipeline 
parallelism, which is the standard tool for bounding communication group size in dense model training~\cite{narayanan2021efficient}, has not been applied to MoE training due to the added complexity of intra-layer expert parallelism.

Piper closes this gap by composing pipeline parallelism with expert parallelism. It organizes $P$ GPUs into a $PP \times EP$ device mesh: $PP$ pipeline stages, each staffed by $EP$ GPUs that handle one partition of experts for $L/PP$ layers via expert-data parallelism. Confining expert-parallel communication to a small, topologically 
local group of GPUs, ideally within a single node or a single-hop Rosetta switch group on Frontier allows Piper to 
exploit fast intra-node interconnects and avoid the high latency of large-scale all-to-all collectives. We expand on Tutel~\cite{tutel2022} to support the underlying expert-parallelism.

Piper consists of four components, described in turn: (i) an analytical resource model that estimates memory, compute, and communication for any $(PP, EP)$ configuration and a model architecture (Section~\ref{sec:memory-model}); (ii) a micro-benchmarking suite that measures platform-specific bandwidth and throughput to parameterize the model (Section~\ref{sec:bench}); (iii) a performance estimator that scores valid configurations by predicted  MFU (Section~\ref{sec:estimator}); and (iv) a pipelined training executor that implements the selected strategy with an efficient 1F1B schedule (Section~\ref{sec:executor}).

\begin{figure*}[!htbp]
    \centering
    \includegraphics[width=0.9\linewidth]{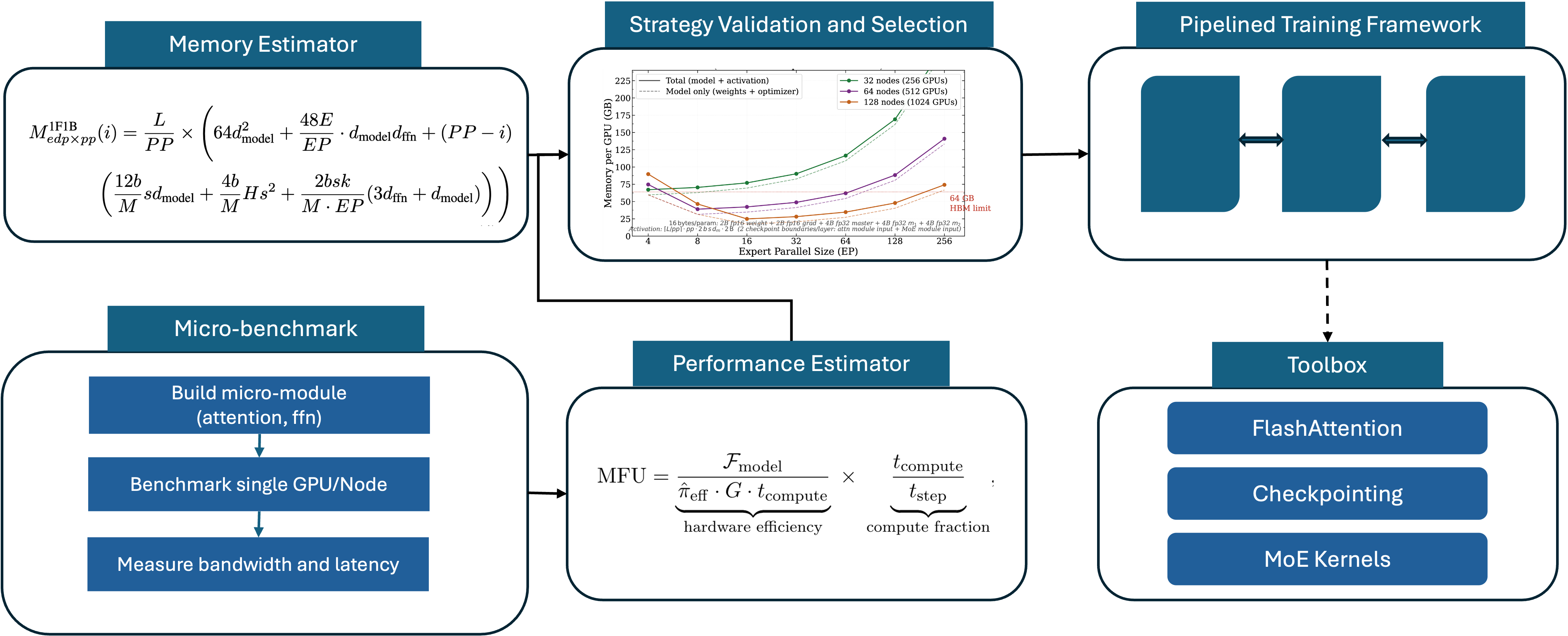}
    \caption{Piper framework for efficient MoE training}
    \label{fig:piper}
\end{figure*}

\begin{figure}[!htbp]
    \centering
    \includegraphics[width=1.0\linewidth]{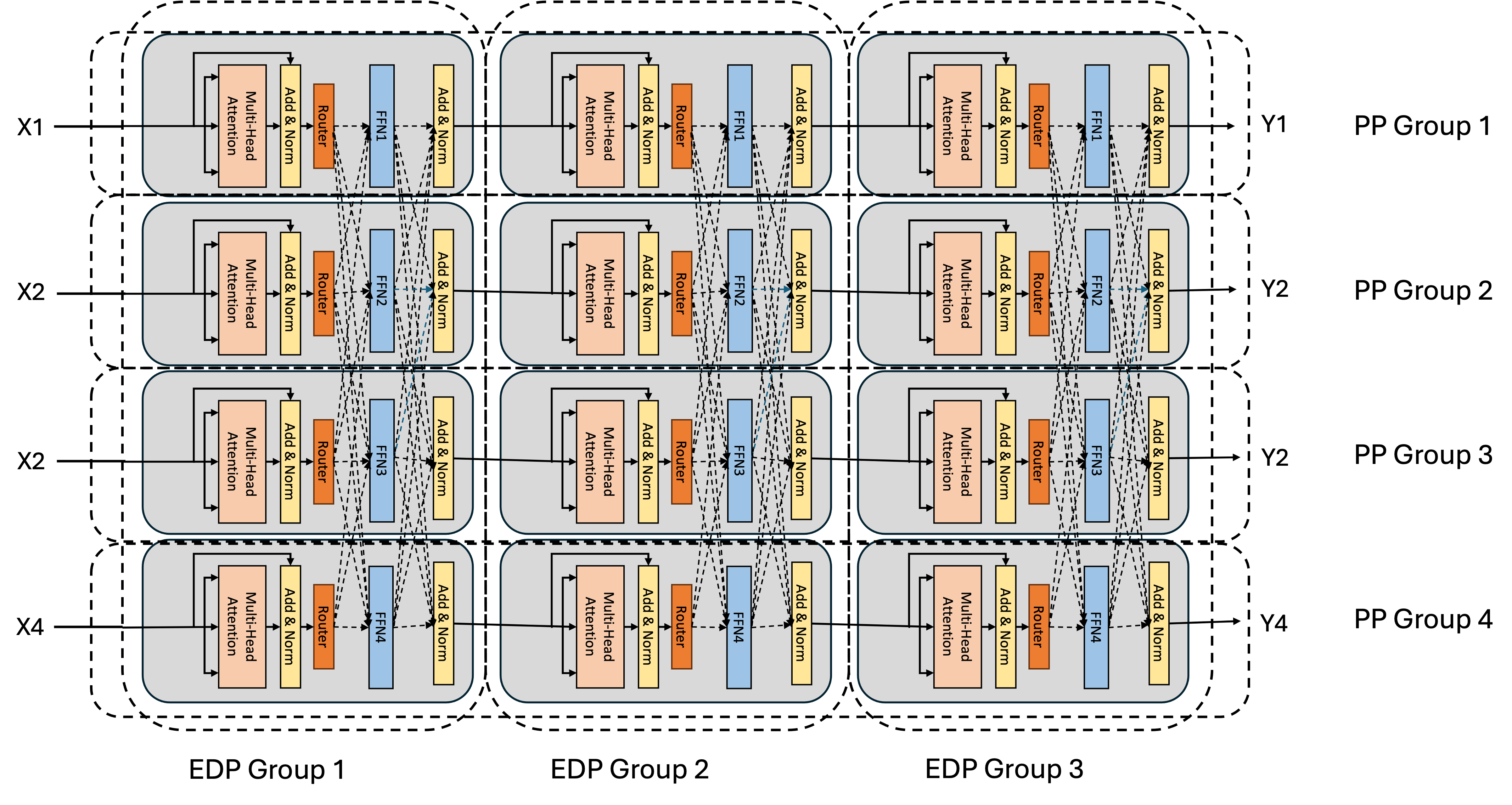}
    \caption{Pipeline Parallelism on Expert Parallelism}
    \label{fig:pp-of-ep}
\end{figure}

At the core of our framework is Pipeline Parallelism on top of expert parallelism (Figure~\ref{fig:pp-of-ep}). We use Tutel for facilitating expert-parallelism. Piper framework has a resource modeling component, a micro-benchmarking suit, a performance estimator, and a pipelined training tool that partitions MoE models across layers.

\subsection{Resource Modeling for MoE Training}
\label{sec:resource-model}

Training large MoE models faces two coupled resource constraints: (i) memory pressure from model parameters, optimizer states, and expert activations, and (ii) communication latency from all-to-all collectives under expert parallelism. We develop analytical models for both, parameterized by the notation in Table~\ref{tab:notation}, and use them to prune the $(PP, EP)$ search space to configurations that are memory-feasible and to rank feasible configurations by predicted throughput.

\paragraph{Notations}

We will use the notations listed in Table~\ref{tab:notation} for resource modeling.

\begin{table}[!htbp]
\centering
\caption{Symbol definitions.}
\label{tab:notation}
\begin{tabular}{cc}
\toprule
\textbf{Symbol} & \textbf{Description} \\
\midrule
$d$                     & Model hidden dimension ($\dmodel$) \\
$L$                     & Total transformer layers (assume all are MoE layers)\\
$L_{\mathrm{MoE}}$      & Number of MoE layers; $L - L_{\mathrm{MoE}}$ are dense \\
$H$                     & Number of attention heads \\
$d_h$                   & Per-head dimension ($H \cdot d_h = d$) \\
$E$                     & Routed experts per MoE layer \\
$E_s$                   & Shared (always-active) experts per MoE layer \\
$k$                     & Top-$k$ routing (experts activated per token) \\
$n_{\mathrm{mat}}$      & Weight matrices/expert: 3 (SwiGLU) \\
$\dffnmoe$              & Expert FFN intermediate dimension \\
$\dffnden$              & Dense FFN intermediate dimension \\
$PP$                    & Pipeline parallel degree \\
$EP$                    & Expert parallel degree \\
$P = PP \times EP$      & Total GPU count \\
$g$                     & GPUs per node \\
$s$                     & Sequence length (tokens) \\
$b$                     & Global batch size (sequences) \\
$b_\mu = b/M$           & Microbatch size \\
$M = \alpha \cdot PP$   & Total microbatches per gradient step \\
$\alpha$                & Microbatch multiplier \\
$i$                     & Pipeline stage index, $0 \le i \le PP-1$ \\
$\betap$                & Bytes per parameter (on GPU) \\
$M_{\mathrm{fw}}$       & Framework overhead (RCCL buffers, etc.) \\
\bottomrule
\end{tabular}
\end{table}

\subsubsection{Modeling Memory}\label{sec:memory-model}
\paragraph{Modeling memory under expert-data parallelism}
The total memory required for training an MoE model constitutes a) static memory (parameter + optimizer states + gradients)  and b) activation memory. 

Mixed-precision training stores parameters in multiple formats simultaneously. The total bytes consumed on GPU per parameter is 16, 2 Bytes for fp16 param, 2 Bytes for fp16 grad,  4 Bytes for fp32 master copy,  and (4 + 4 = 8) Bytes fp32 momentum and variance.

We first establish a lower bound on the memory by assuming everything fits in a hypothetical GPU with infinite memory so that no model parallelism is required (Table~\ref{tab:undivided-memory}). There are $4d_{model}^2$ attention parameters ($W_K, W_Q, W_V \in \mathbf{R}^{d_{model}\times d_{model}}$) for Multi-Head Attention (MHA). Each of the FFN experts has $3d_{model}d_{ffn}$ weights since there are three weight matrices ($W_{up}, W_{gate} \in \mathbf{R}^{d_{model}\times d_{ffn}}$ and $W_{down} \in \mathbf{R}^{d_{ffn}\times d_{model}})$. 

Accounting for the activation memory, intermediate output is calculated in half precision (2 Bytes), for $b$ sequences with $s$ sequence length, each expert receives $s_e; E[s_e] \approx \dfrac{bsk}{E}$ tokens. For a single expert with SwiGLU activations, each token activating an expert creates $3d_{ffn}$ (with fused kernel) or $4d_{ffn}$ values (up, gate, down). Since $E[s_e] = bsk/E$, per expert activation memory is $2 Bytes \times bsk/E \times (3d_{ffn} + d_{model})$. Activation memory from the attention module is $2 Bytes \times (6bsd_{model}$ (Q, K, V projections, Attention output, output projections) + $2 Bytes \times 2bHs^2$ (Attention score, softmax output) (Table~\ref{tab:undivided-memory}). With, flash attention, $4bHs^2 \rightarrow 2bHs$.

\begin{table}[!htbp]
    \centering
    \begin{tabular}{cccc}
            \toprule
            Model & \#Parameters & Model Memory & Activation Memory \\
            \midrule
            Attention & $4d_{model}^2$ & $64d_{model}^2$ & $12bs\, d_{model} + 4bHs^2$ \\
            \midrule
            Experts & $3Ed_{model} d_{ffn}$ & $48 Ed_{model} d_{ffn}$ & $2bs k (3d_{ffn} + d_{model})$ \\
            \bottomrule
        \end{tabular}
    \caption{Estimating training memory}
    \label{tab:undivided-memory}
\end{table}

\begin{equation}
\begin{split}
    M_u = L\Big(64 d_{model}^2\ +\ 48 Ed_{model}d_{ffn}\ +\ 12bsd_{model} \\ 
    +\ 4Hbs^2\ +\ 2bsk\big(3d_{ffn}\ +\ d_{model}\big)\Big)
    \label{eq:undivided_memory}
\end{split}
\end{equation}

\subsubsection{Memory with Expert Data Parallelism}
In expert data parallelism, the world size is $P=EP$. The non-expert modules (attention) get replicated across $P$ GPUs and each GPU gets $E/EP$ experts. The total memory consumption increases due to the replicated attention module. However, the per GPU memory requirement comes down to

\begin{equation}
\begin{split}
    M_{edp} = L\Big(64 d_{model}^2 + \dfrac{48 E}{EP}d_{model}d_{ffn} + 12 bsd_{model} \\
    +\ 4Hbs^2 + \dfrac{2bsk}{EP}\big(3d_{ffn} +d_{model}\big)\Big)
\end{split}
\end{equation}

\paragraph{Memory Under Pipelined Expert Parallelism}

In the hybrid Pipeline-Expert parallelism, we divide the $P$ GPUs into $PP\times EP$ grid and there are $EP$ pipeline parallel groups of size $PP$ and there are $PP$ expert-parallel groups of size $EP$. In each of the $PP$ pipeline stages, there are $EP$ GPUs hosting $l=L/PP$ layers. Each of these $EP$ GPUs is replicating the non-expert parameters (attention, router) and hosting $E/EP$ experts in an expert-data parallel way. So, each GPU's memory is $l \times (attention\text{-}memory + E/EP\ expert\text{-}memory)$. The peak activation memory in GPipe schedule is for all of the $M$ microbatches since at the steady state all microbatches needs to stay alive in the memory.

\begin{equation}
    \begin{split}
        M_{edp\times pp}^{GPipe} = \dfrac{L}{PP}\Big(64 d_{model}^2\ + \dfrac{48 E}{EP}d_{model}d_{ffn} \\
    +\ 12bsd_{model}\ +\ 4Hbs^2\ +\ \dfrac{2bsk}{EP}\big(3d_{ffn}\ +\ d_{model}\big)\Big)
    \end{split}
\end{equation}

\paragraph{1F1B Pipeline Schedule}

In 1F1B scheduling, stage $i$ (0-indexed, $i \in \{0, 1, \ldots, PP-1\}$) holds 
$(PP - i)$ in-flight microbatch activations simultaneously at peak. Each microbatch has size $b/M$, where $M$ is the total number of microbatches. Hence, the per-GPU memory for stage $i$ is:

\begin{equation}
    \begin{split}
    M_{edp\times pp}^{\text{1F1B}}(i) = \dfrac{L}{PP} \times \Bigg( 64d_{\text{model}}^2 + \frac{48E}{EP} \cdot d_{\text{model}} d_{\text{ffn}} 
        + (PP - i) \\ \left(\frac{12b}{M} s d_{\text{model}} + \frac{4b}{M} H s^2 + \frac{2bsk}{M \cdot EP}(3d_{\text{ffn}} + d_{\text{model}})\right) \Bigg)
    \end{split}
\end{equation}

\begin{comment}
\subsubsection{Boundary Cases}

At \textbf{stage 0} (maximum memory):

\begin{equation}
    M_{\text{1F1B}}(0) = \frac{X}{PP} \times \Bigg(
        48d_{\text{model}}^2 
        + \frac{36E}{EP} \cdot d_{\text{model}} d_{\text{ffn}} 
        + PP \left(
            \frac{12b}{M} s d_{\text{model}} 
            + \frac{4b}{M} H s^2 
            + \frac{2bsk}{M \cdot EP}(3d_{\text{ffn}} + d_{\text{model}})
        \right)
    \Bigg)
\end{equation}

At \textbf{stage $PP - 1$} (minimum memory):

\begin{equation}
    M_{\text{1F1B}}(PP-1) = \frac{XYZ}{PP} \times \Bigg(
        48d_{\text{model}}^2 
        + \frac{36E}{EP} \cdot d_{\text{model}} d_{\text{ffn}} 
        + 1 \cdot \left(
            \frac{12b}{M} s d_{\text{model}} 
            + \frac{4b}{M} H s^2 
            + \frac{2bsk}{M \cdot EP}(3d_{\text{ffn}} + d_{\text{model}})
        \right)
    \Bigg)
\end{equation}

\end{comment}
%\subsubsection{Memory Imbalance Across Stages}

The memory difference between the first and last stage is:

\begin{equation}
    \begin{split}
    \Delta M &= M_{\text{1F1B}}(0) - M_{\text{1F1B}}(PP-1) \\
    &= \frac{L(PP-1)}{PP} \left(
        \frac{12b}{M} s d_{\text{model}} 
        + \frac{4b}{M} H s^2 
        + \frac{2bsk}{M \cdot EP}(3d_{\text{ffn}} + d_{\text{model}})
    \right)
    \end{split}
\end{equation}

That means, the first stage needs to hold $(PP-1)$-times more activation memory than the last stage. This creates heavily skewed memory pressure across pipeline stages.

%\clearpage

\subsection{Modeling Communication}
\subsubsection{Communication Under Expert-Data Parallelism}

Under expert data parallelism, major communication happens due to the activation values. There are two phases of all-to-all communications.

\paragraph{Dispatch and Combine}
Each data-parallel router routes it's $bsk$ tokens ($bs$ tokens, but each token gets to $k$ experts) to $E$ experts. Under proper load-balancing, every GPU sends $bsk/EP$ tokens to every other GPU. The communication volume between a pair of GPUs is  $2bsk/EP$ and total communication volume during the dispatch all-to-all is $\binom{EP}{2} \times \dfrac{2bsk}{EP} = (EP-1)bsk$. These are tokens, so they have $d_{model}$ dimensions, in fp16/bf16, the total message size is $2(EP-1)bsk \times d_{model}$ bytes. Individual message size is $\approx \dfrac{4bsk d_{model}}{EP}$  bytes.

Combine is the same communication in the reverse direction. So, the message volumes are same. The per-GPU send volume during dispatch is $\frac{2bskd_{\mathrm{model}}}{EP}$ 
bytes (fp16), giving a per-NIC injection load of $\frac{4bskd_{\mathrm{model}}}{EP \cdot B_{\mathrm{NIC}}}$ seconds at NIC bandwidth $B_{\mathrm{NIC}}$. Since combine is the reverse operation, the total all-to-all latency per MoE layer in the forward pass is bounded by:

\begin{equation}
    T_{\mathrm{a2a}} \geq \frac{4bskd_{\mathrm{model}}}{EP \cdot B_{\mathrm{NIC}}}
    \label{eq:a2a-latency}
\end{equation}

This bound is tight when NICs are uniformly saturated; Section~\ref{sec:alltoall} 
discusses why the flat RCCL all-to-all fails to achieve it on 
Dragonfly topologies and how our topology-aware algorithm addresses this.

\subsubsection{Communication under Pipeline and Expert-Data Parallelism}
World size $P = EP \times PP$. For each stage hosting $l$-layers, there are $2l$ all-to-all communication calls in the forward pass. 

Between tow stages a batch of P2P communication happens. Each of the $EP$ GPUs hosting the last layer of the $i^{th}$ stage sends $2bsd_{model}$ bytes to it's counterpart to the first layer of the $(i + 1)^{st}$ stage. Total message sent between two stages in this stage is $2EP\times bsd_{model}$ bytes. Now, under different pipeline scheduling, number of stages participating in this concurrent communication varies.

%We consider three pipeline schedules, namely GPipe, 1F1B, and 1F1B interleaved. 

\subsection{Finding Valid Parallelization Strategies w/o OOM}
\begin{align}
  PP \times EP &= n \times g  &&\text{(total GPU count)} \\
  EP &\mid E                  &&\text{(EP divides expert count)} \\
  PP &\leq L                  &&\text{(} \geq 1 \text{ layer per stage)} \\
  EP &\leq g \cdot N_h        &&\text{(EP within fast-interconnect domain)} \\
  M_{\mathrm{peak}}(0) &\leq C_{\mathrm{GPU}} &&\text{(worst-case stage fits in HBM)}
\end{align}

where $N_h$ is the number of nodes sharing a single-hop interconnect (e.g., $N_h = 4$ for a Rosetta switch group on Frontier). The fourth constraint ensures that all-to-all communication during expert dispatch stays within the fast intra-group fabric. The fifth uses the stage-0 peak from the 1F1B model, which is the binding memory constraint across all stages.

\subsection{Pipelined Training Execution}
Among valid configurations, Piper ranks them by estimated MFU. Then it interfaces PyTorch Distributed Pipeline Parallelism with expert-parallel or any other hybrid intra-node distribution library such as Tutel. We instrumented 1F1B schedule and installed synchronization mechanisms among expert-parallel group members. We expanded Tutel and PyTorch's pipeline parallelism mechanism to work with each other under a two-dimensional parallelization (three, counting external data parallelism). 

\begin{comment}
\begin{equation}
    \mathrm{MFU} = \frac{C_{\mathrm{model}}}{T_{\mathrm{step}} 
    \cdot P \cdot \pi_{\mathrm{GPU}}}
\end{equation}

where $C_{\mathrm{model}}$ is the theoretical FLOPs per step (Section~\ref{sec:flop-model}), $T_{\mathrm{step}}$ is the estimated step time from the performance estimator, and $\pi_{\mathrm{GPU}}$ is the peak FP16 throughput of a single GPU.

Using the memory model described in Section~\ref{sec:memory-model}, we develop a tool to find viable parallelization strategies and minimum number of nodes required to train a model. A configuration $(PP, EP)$ on $n$ nodes with $g$ GPUs per node is valid if and only if all of the following hold:
 
\begin{align}
  PP \times EP &= n \times g
  & &\text{(total GPU count)} \\
  EP &\mid E
  & &\text{(EP divides number of experts)} \\
  PP &\leq L
  & &\text{(each stage holds} \geq 1 \text{ layer)} \\
  M_{\mathrm{peak}} &\leq C_{\mathrm{GPU}}
  & &\text{(fits in HBM)}
\end{align}
\end{comment}

\section{Performance Modeling through Micro-benchmarking}\label{sec:mfu-estimator}
Resource modeling gives us a realistic expectation regarding viable distributed training strategies in terms of number of nodes, degrees of expert and pipeline data parallelism, whether to adopt memory saving techniques such as checkpoint activation, offloading, etc. This can be done largely statically by the mathematical formulas we developed in Section~\ref{sec:resource-model} and considering system properties such as HMB memory. Once we find candidate strategies to train our model without running into out of resource (e.g., OOM) error.

To estimate the model flops utilization or MFU for each of these strategies, we need to micro-benchmark the HPC platform for various phases of the MoE training workflow. With pipelined expert-data parallelism, the typical flow is
{\small
\[
\Big(attn \rightarrow routing \rightarrow dispatch\_a2a \rightarrow expert \rightarrow combine\_a2a\Big)_{\times l} \rightarrow P2P
\]
}

So, we need to run micro-benchmarking on computational performance and communication latency. 

\subsection{Micro-benchmarking Computation}\label{sec:bench}

Every transformer layer has two major components, non-expert attention module, and multiple Feed Forward Network (FFN) experts. In an expert-data parallel setting, we need to distribute the experts equally across the $g$ gpus in a node and replicate the attention module in a data-parallel way across all $g$ gpus. Each GPU needs to host and process one attention module and $E/g$ experts. We start with separate optimization of these two parts and identify the suitable kernels optimized for each part. 

\subsubsection{Attention Performance}
\begin{figure}[!htbp]
    \centering
    \includegraphics[width=0.9\linewidth]{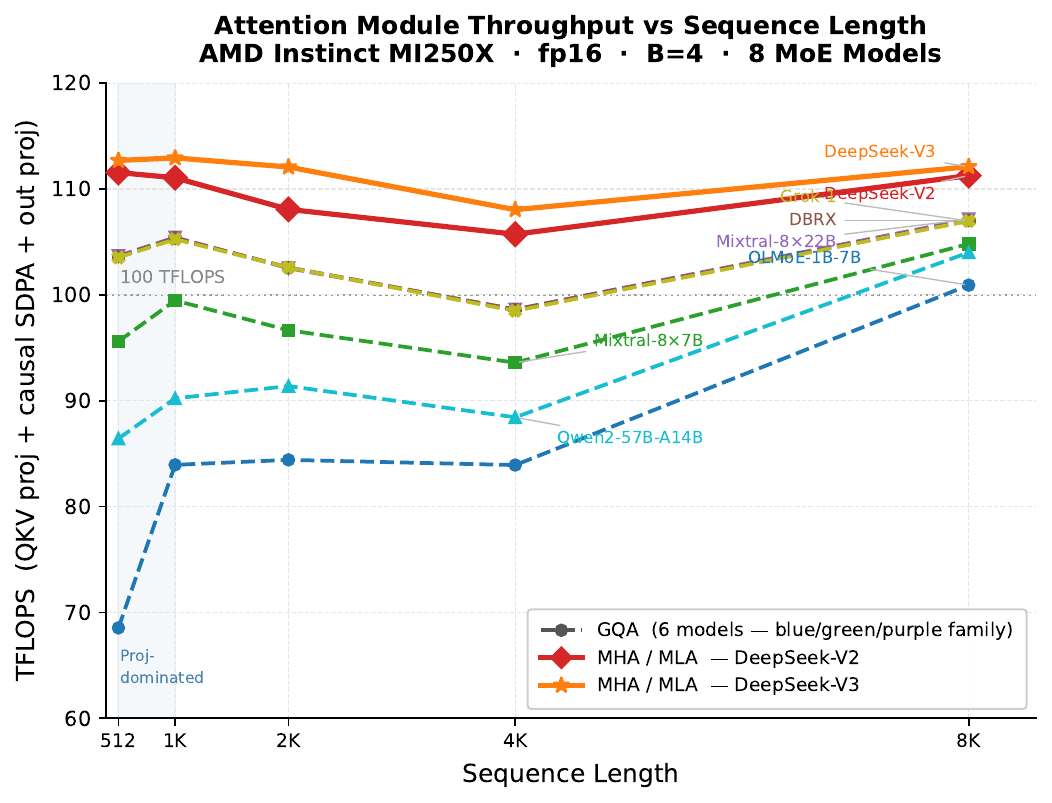}
    \caption{Achievable throughput for various models for different sequence lengths.}
    \label{fig:attn-bench}
\end{figure}

For attention part, different models have different model dimension, and the flash-attention kernel is optimized for only a fixed set of head-dimension. We benchmark a single GPU to identify the best performing head-dimension. We need to choose a set of best throughput generation $<batch\_size, head\_dimension>$ tuples. Figure~\ref{fig:attn-bench} shows our measure performances for various MoE model architectures.

\subsubsection{Expert Performance}
\begin{figure}[!htbp]
    \centering
        \includegraphics[width=0.9\linewidth]{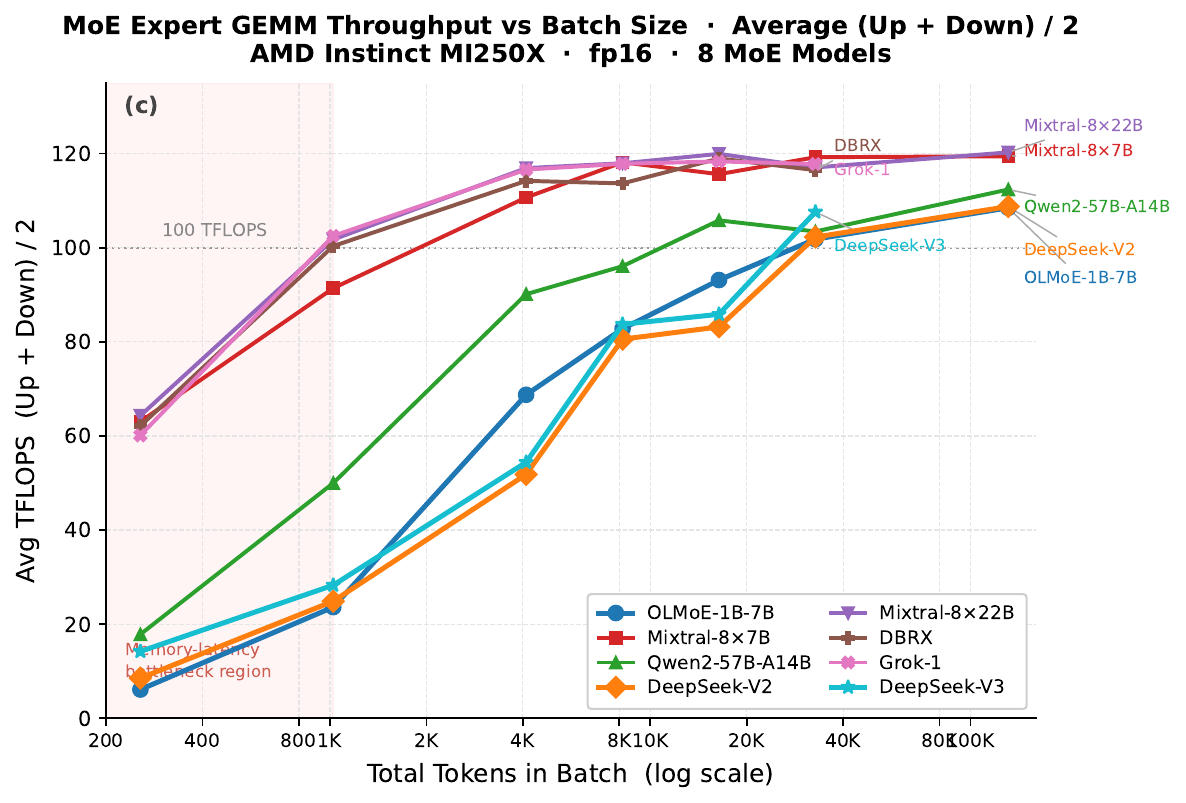}
        \caption{MoE GEMM Performance}
    \label{fig:expert-bench}
\end{figure}

Similarly, for the FFN part, which mostly requires GEMM operations, we need to identify the set of (num\_tokens, batch\_size, expert\_dimension) 
tuples that gives us the best throughput. For the fine-grained experts where GEMM between many tall and skinny matrices are involved, the search becomes more involved (Figure~\ref{fig:expert-bench}).

Then, from these two sets, we need to identify the tuples with agreeing $batch\_size$. Choosing this {batch\_size} becomes a critical choice, especially in the context of pipeline parallelism since choosing the right number of micro batch size, and right value for micro batch size will determine pipeline bubble and computation efficiency.

\subsection{Micro-benchmarking Communication}
In our framework, we incur two types of communication, all-to-all within an expert-parallel group and send-recv between two pipeline stages. Since, between stage send-recv communication happens following a synchronization across the expert parallel group, there are EP concurrent P2P communications between two stages.

\begin{figure}[!htbp]
    \centering
    \includegraphics[width=0.5\textwidth]{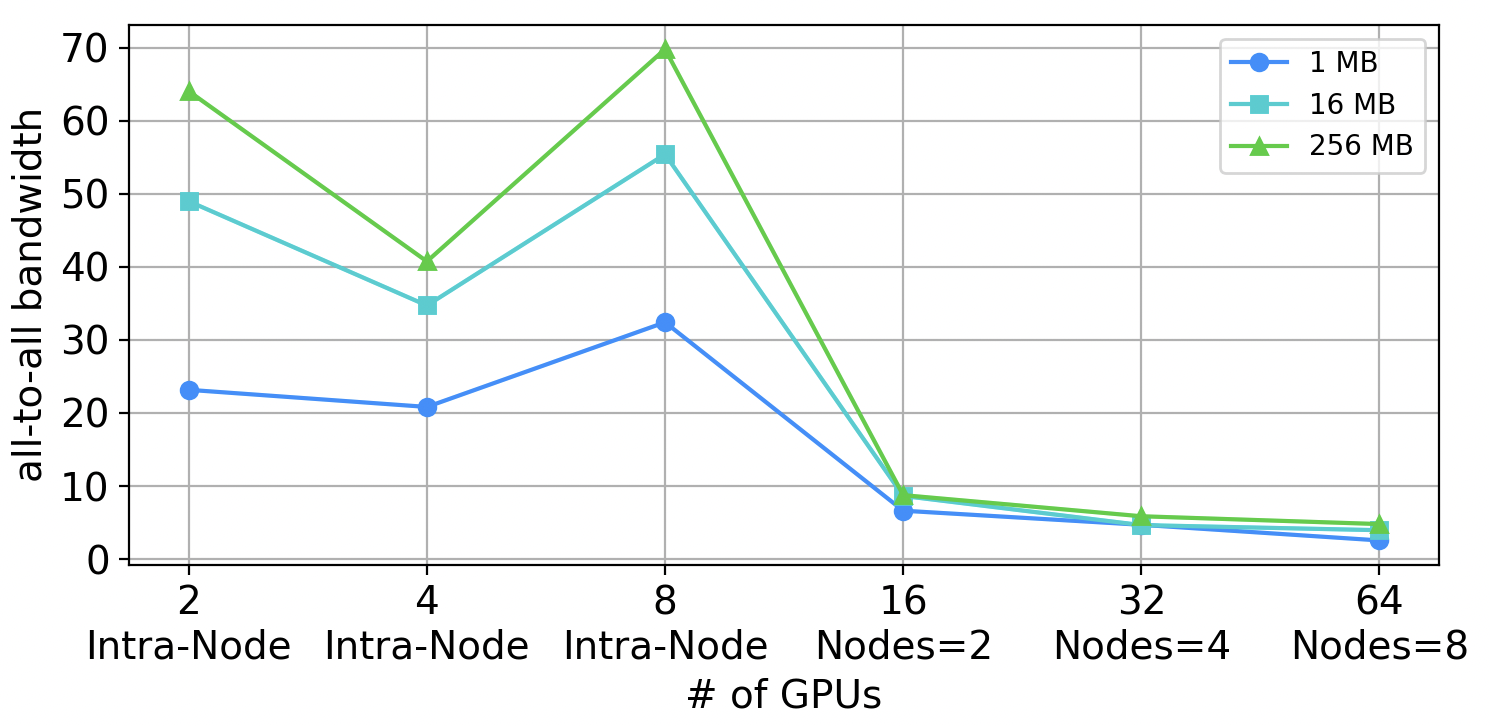}
    \caption{Benchmarking All-to-all bandwidth for various message sizes. Bandwidth drops significantly as soon as all-to-all involves inter-node communicaiton}
    \label{fig:a2a-bench}
\end{figure}

For benchmarking all-to-all bandwidth, we vary the number of GPUs from 2 to 64 spanning 1-8 nodes, for various message sizes (Figure~\ref{fig:a2a-bench}).

\subsection{MFU Estimation}\label{sec:estimator}
Let $t_{\mathrm{compute}}$ and $t_{\mathrm{comm}}$ be the attention + expert and A2A + P2P
totals summed over the full step. Then:

\begin{equation}
\mathrm{MFU}
  = \underbrace{
      \frac{\mathcal{F}_{\mathrm{model}}}{\hat{\pi}_{\mathrm{eff}}\cdot G \cdot t_{\mathrm{compute}}}
    }_{\text{hardware efficiency}}
  \times
  \underbrace{
      \frac{t_{\mathrm{compute}}}{t_{\mathrm{step}}}
    }_{\text{compute fraction}},
\qquad
\frac{t_{\mathrm{compute}}}{t_{\mathrm{step}}}
  = 1 - b - \frac{t_{\mathrm{comm}}}{t_{\mathrm{step}}}
\end{equation}
\section{HALO: Hierarchical Affinity-aware Locality-Optimized All-to-All}\label{sec:alltoall}

Our algorithm pursues five design goals. It saturates all four NICs on a single node during inter-node communication, and identifies the three phases of communication along with their dependency structure in order to maximize concurrency across phases. At the intra-node level, it enforces maximum locality by exploiting GPU-to-NIC affinity. At the inter-node level, it leverages the Dragonfly grouping structure by treating the four nodes connected to a common Rosetta switch as a single communication locality domain. Finally, when possible, it constrains node allocation to within the same rack entirely, avoiding the slowest inter-rack communication links.

\subsection{Communication Group Construction}

We assign GPUs sharing a NIC to the same inter-node communicator, ensuring that inter-node traffic from a given GPU is always injected through its affinitized NIC. This prevents NIC contention and allows all four NICs on a node to operate at full bandwidth simultaneously.
\begin{figure}[h]
    \centering
    \includegraphics[width=1.0\linewidth]{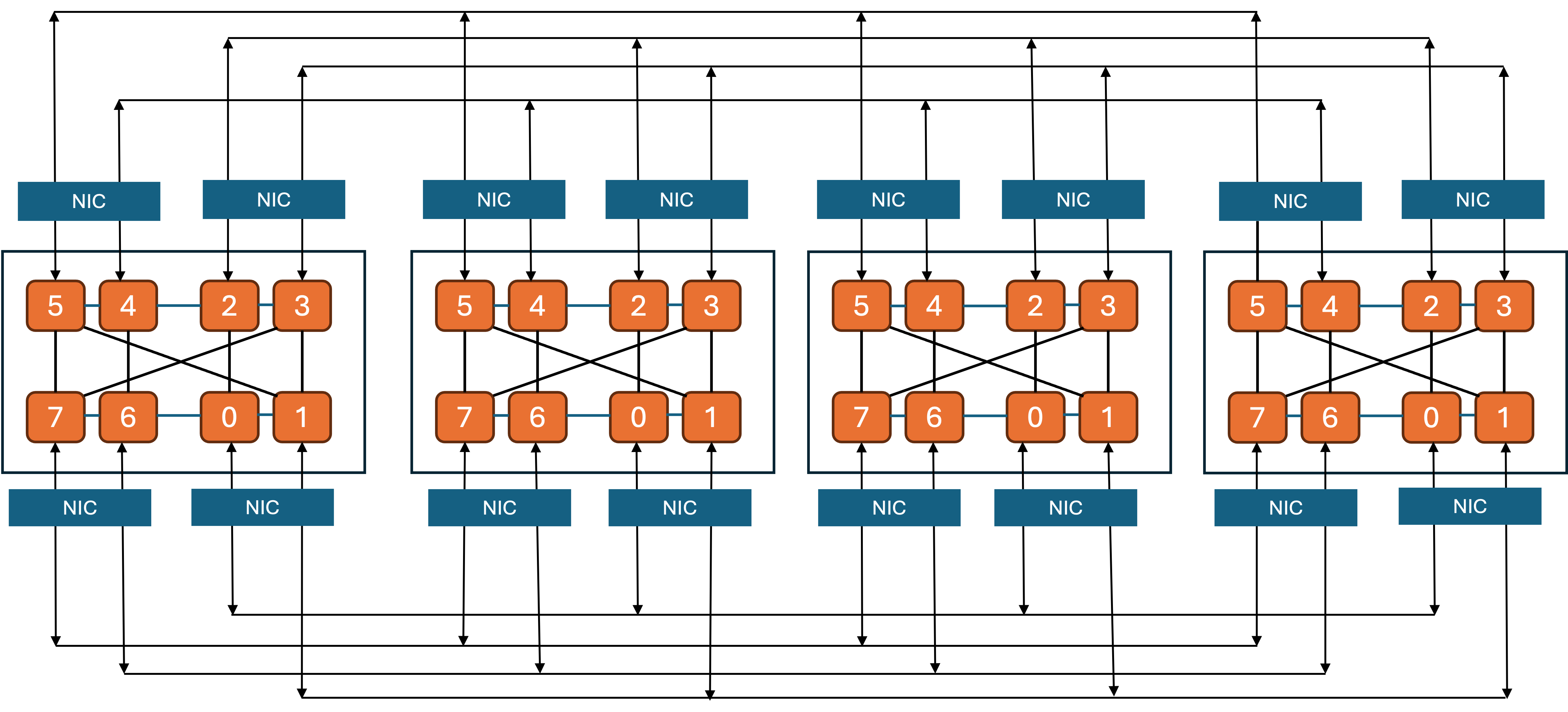}
    \caption{Communication groups for Topological all-to-all}
    \label{fig:topo-a2a}
\end{figure}

\subsection{Three-Phase Algorithm and Dependency Structure}
% ============================================================
%  Algorithm 2 — Main All-to-All  (spans both columns)
% ============================================================
\begin{algorithm}[!htbp]
\small
\DontPrintSemicolon
\caption{\textsc{HALO AllToAll}%
  %$\bigl(\mathbf{S},\;\mathit{int\_grp},\;\mathit{nic\_grp},\;
   %\mathit{local\_rank},\;\mathit{node\_id},\;\mathit{num\_nodes},\;
   %\mathit{nic\_rank},\;\mathit{device}\bigr)$
   }
\KwIn{$\mathbf{S}\!\in\!\mathbb{R}^{N\times D}$ send buffer, $N = \mathit{num\_nodes}{\times}R$ \newline communication groups and rank metadata from \textsc{SetupComms}}

\KwOut{$\mathbf{O}\!\in\!\mathbf{R}^{N\times D}$ fully transposed output buffer}

\BlankLine
$\mathit{local\_start} \leftarrow \mathit{node\_id}\cdot R$;\quad
$\mathcal{N}_{-} \leftarrow \{n\in[0,\mathit{num\_nodes}) : n\neq\mathit{node\_id}\}$;\quad
$M \leftarrow |\mathcal{N}_{-}|$\;

\BlankLine
\tcp{Persistent buffers — allocated once on first call, reused thereafter}
$\mathbf{P}_1\!\in\!\mathbb{R}^{R\times D}$\quad\mycomment{Ph1 recv}\;
$\mathbf{F}^{s}_{2},\;\mathbf{F}^{r}_{2}\!\in\!\mathbb{R}^{MR\times D}$\quad\mycomment{Ph2 send/recv flat buffers}\;
$\mathbf{F}^{s}_{3},\;\mathbf{F}^{r}_{3}\!\in\!\mathbb{R}^{R\times MD}$\quad\mycomment{Ph3 send/recv flat buffers}\;
$\mathbf{O}\!\in\!\mathbb{R}^{N\times D}$\mycomment{output}\;

\BlankLine
%% ---- Phase 1 ----
\PhaseOne{
  \tcp{Contiguous copy of this node's $R$ rows}
  $\mathbf{L} \leftarrow \mathbf{S}[\mathit{local\_start}:\mathit{local\_start}{+}R]$
    \quad\mycomment{shape \sh{R,D}}\;
  \BlankLine
  \tcp{Intra-node exchange: rank $\ell$ receives row $\ell$ from every local peer}
  $\mathbf{P}_1 \leftarrow \textsc{AllToAllSingle}\!\left(\mathbf{L},\;\mathit{internal\_group}\right)$\;
}

\BlankLine
%% ---- Phase 2 ----
\PhaseTwo{
  \tcp{Pack all remote-destined rows with a single \textsc{IndexSelect} kernel}
  $\mathit{idx} \leftarrow \bigl[nR,\ldots,nR{+}R{-}1\;\big|\;n\in\mathcal{N}_{-}\bigr]$
    \mycomment{pre-built LongTensor on GPU}\;
  $\mathbf{F}^{s}_{2} \leftarrow \textsc{IndexSelect}\!\left(\mathbf{S},\;\dim=0,\;\mathit{idx}\right)$
    \mycomment{one GPU kernel; shape \sh{MR,D}}\;
  \BlankLine
  \tcp{Batched P2P over world group (RDMA); one send+recv slice per remote node}
  \For{$i,\;n \in \textsc{Enumerate}(\mathcal{N}_{-})$}{
    $\mathit{peer} \leftarrow n\cdot R + \mathit{local\_rank}$\;
    $\textsc{IRecv}\!\left(\mathbf{F}^{r}_{2}[iR:(i{+}1)R],\;\mathrm{src}=\mathit{peer}\right)$\;
    $\textsc{ISend}\!\left(\mathbf{F}^{s}_{2}[iR:(i{+}1)R],\;\mathrm{dst}=\mathit{peer}\right)$\;
  }
  $\textsc{WaitAll}()$\quad\mycomment{flush all RDMA ops}\;
}

\BlankLine
%% ---- Phase 3 ----
\PhaseThree{
  \tcp{Transpose Ph2 recv buffer; one GPU kernel — no intermediate allocation}
  $\mathbf{F}^{s}_{3} \leftarrow
    \textsc{Reshape}\Bigl(
      \textsc{Permute}\!\bigl(
        \textsc{View}(\mathbf{F}^{r}_{2},\;M,R,D),\;
        [1,0,2]
      \bigr),\;
      [R,\;MD]
    \Bigr)$
    \quad\mycomment{row $\ell$ = all remote data for local rank $\ell$}\;
  \BlankLine
  \tcp{Scatter repacked rows to final owners on this node}
  $\mathbf{F}^{r}_{3} \leftarrow
    \textsc{AllToAllSingle}\!\left(\mathbf{F}^{s}_{3},\;\mathit{internal\_group}\right)$
    \quad\mycomment{rank $\ell$ receives its $\sh{MD}$-wide slice}\;
}

\BlankLine
%% ---- Combine ----
\tcp{Write Phase 1 (local) and Phase 3 (remote) results into output buffer}
$\mathbf{O}[\mathit{local\_start}:\mathit{local\_start}{+}R] \leftarrow \mathbf{P}_1$\;
$\mathbf{T} \leftarrow
  \textsc{Permute}\!\bigl(\textsc{View}(\mathbf{F}^{r}_{3},\;R,M,D),\;[1,0,2]\bigr)$
  \quad\mycomment{shape \sh{M,R,D}}\;
\For{$i,\;n \in \textsc{Enumerate}(\mathcal{N}_{-})$}{
  $\mathbf{O}[nR:(n{+}1)R] \leftarrow \mathbf{T}[i]$\;
}

\BlankLine
\Return $\mathbf{O}$\;
\end{algorithm}

\begin{figure}[!htbp]
    \centering
    \includegraphics[width=0.9\columnwidth]{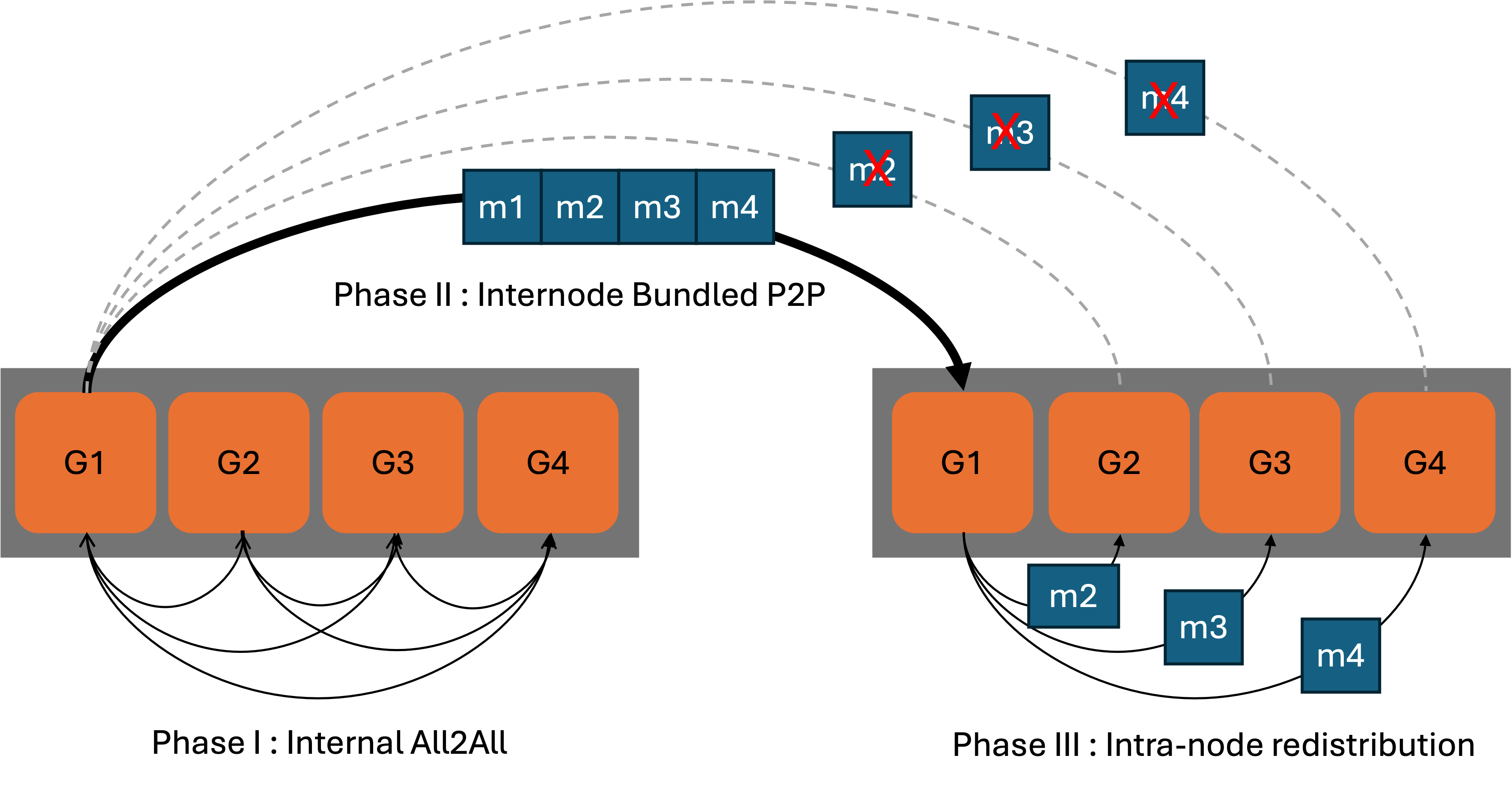}
    \caption{Three phases of our HALO all-to-all algorithm.  Phase I can happen concurrently with either of Phase II and Phase III, but Phase III needs to wait for Phase II to complete.}
    \label{fig:halo-phases}
\end{figure}

The three phases exhibit the following dependency structure:
\begin{equation}
    \text{Phase~I} \;\|\; \bigl(\text{Phase~II} \;\rightarrow\; \text{Phase~III}\bigr)
\end{equation}
Phase~I (intra-node all-to-all) is fully independent of the inter-node phases, since all source-destination pairs are known locally and  require no prior accumulation. Phase~III depends on the completion of  Phase~II, as it redistributes the received remote data within the node. We exploit the Phase~I independence by launching it concurrently with  Phase~II or Phase~III, hiding a significant portion of intra-node communication  latency behind inter-node transfers. Figure~\ref{fig:halo-phases} demonstrates our three phases and their overlapping strategies.

\subsection{Rack-Aware Node Allocation}
% This is currently missing but was promised in your intro paragraph

When the job scheduler permits, HALO constrains node allocation to  within a single rack to eliminate inter-rack traffic, which traverses  the slowest links in the Dragonfly topology. Within a rack, nodes are  further grouped by Rosetta switch affinity: the four nodes sharing a 
switch form a single locality domain, and inter-group communication is batched across this boundary to minimize contention on inter-switch links.

\subsection{Comparison against torch.dist.all\_to\_all}

\begin{figure}[!htbp]
    \centering
    \includegraphics[width=1.0\linewidth]{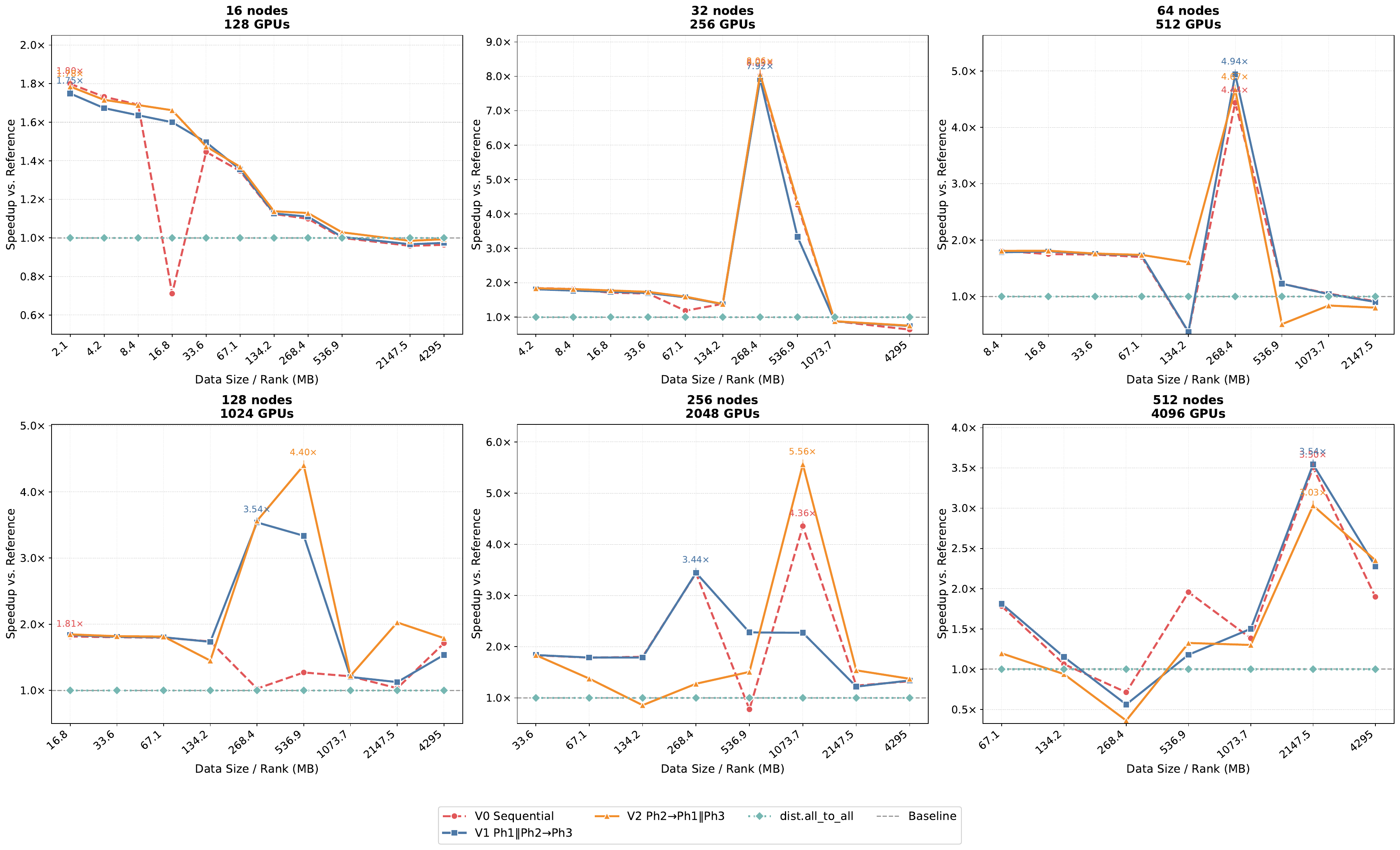}
    \caption{Latency comparison of Neighborhood-all\_to\_all algorithm against the RCCL based torch.dist.all\_to\_all algorithm.}
    \label{fig:a2a_speedup}
\end{figure}

We compare HALO against \texttt{torch.dist.all\_to\_all} backed by RCCL across varying node counts and message sizes (Figure~\ref{fig:a2a_speedup}). HALO achieves 1.1$\times$--9$\times$ 
lower latency for configurations of 16 nodes or more. The crossover at 16 nodes reflects the threshold at which inter-rack communication becomes dominant under the flat RCCL implementation — precisely the regime where HALO's rack-aware grouping and NIC saturation strategy yield the largest gains. At smaller scales, where all GPUs fit within a single switch group, both algorithms perform comparably since single-hop communication already saturates available bandwidth.

\section{Load Balancing through Expert Migration}\label{sec:migration}

The load across GPUs emerging from router's expert selection can vary over training period. Initially, all experts are preferred by the router equally; however small random difference can make some experts more favorable. These favored experts get more tokens, and consequently more gradient updates. This creates a positive feedback loop where a subset of experts become more capable than others, and end up getting more tokens consistently. This causes expert collapse.

In the middle part of the training, experts become specialized in different types of input. The load distribution stabilizes, and different experts develop distinct activation patterns across the feature space. Towards the end, the specialization becomes more concrete. Some experts remain underutilized.

\begin{figure}[!htbp]
    \centering
    \includegraphics[width=1.0\linewidth]{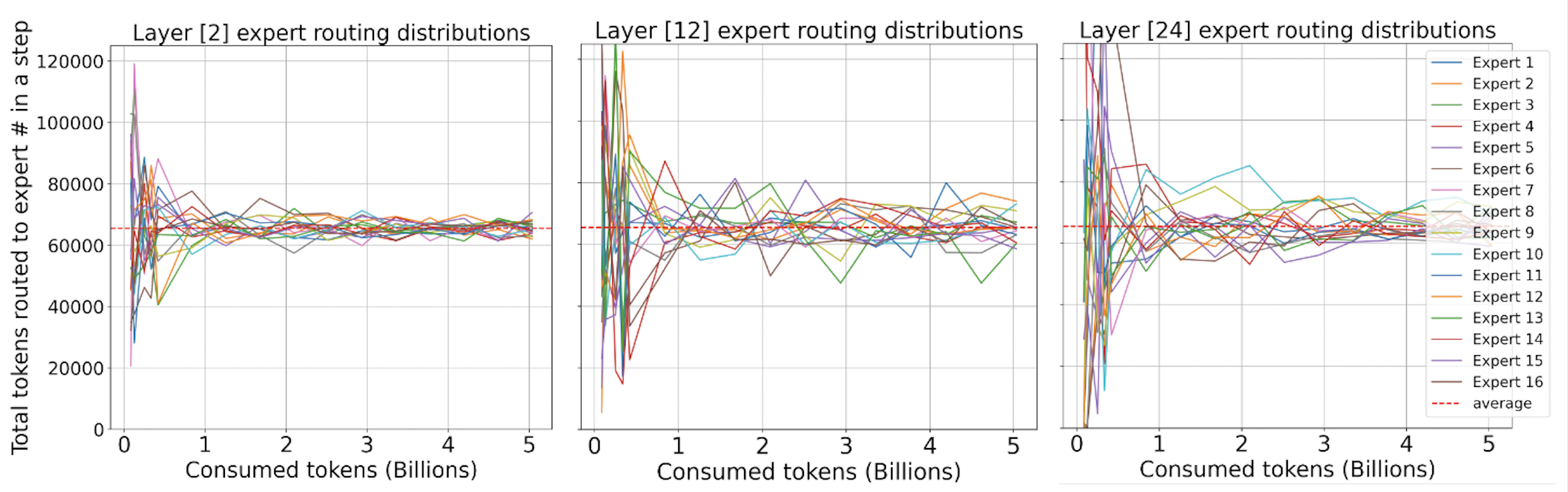}
    %\vspace{-10pt}
    \caption{Expert load distribution during the training process. The model has a 350M base model, with 16 experts and expert frequency 0.5.}
    \label{fig:frequency}
\end{figure}

Unless there is an auxiliary load-balancing loss or other load balancing mechanism in place, expert workload remains skewed. So, a static expert parallelism at the beginning of the training cannot ensure that all the GPUs within a layer are receiving balanced workload. While load balancing loss helps with load balancing in the middle stages of the training, it still suffers from prolonged period of load imbalance. Figure~\ref{fig:frequency} shows expert-routing distribution in different layer becomes balanced after almost one billion token consumption. 

Existing MoE training methods are averse to expert re-assignment for load balancing due to the perceived overhead of expert migration. Under existing distribution techniques that distribute experts among many nodes through slow Ethernet, this caution is justified. However, when we successfully utilize pipeline parallelism to localize the experts from the same layer to a small number of GPUs connected via fast link (NVLink or Infinity Fabric), we should reconsider the dynamic load balancing. In the absence of load balancing loss, the expert demands are still skewed, but load across the GPUs can be balanced. Even with the expert collapse, the final model performance does not suffer.

\subsection{Dynamic Expert Migration}
We extend the router class to maintain token distribution throughout the training. An external scheduler (could be intermittent interrupt in the training loop) inspects the growing load imbalance, and whenever it crosses a pre-determined threshold, it can run our migration algorithm to identify the minimal number of intra-group (expert-parallel group) migrations of experts among GPUs. The cost of migration will be the cost of an all-to-all in the expert parallel group where individual expert's migration cost is the latency of moving $48 d_{model} d_{ffn}$ ($3 d_{model} d_{ffn}$ expert parameters, and each parameter needs 16 Bytes for master copy, optimizer states, and gradients). In the worst case, when we do a complete re-assignment from scratch, that would mean we will be moving all $E$ experts, and the individual send message size will be $48 (E/EP)d_{model} d_{ffn}$ ($3 d_{model} d_{ffn}$ Bytes. Assuming a single layer fits in a node (8 GPUs) with EP=8, the individual send-message sizes for various SOTA model is in Table~\ref{tab:migration-cost}.

\begin{table}[!htbp]
\centering
\caption{Per-Layer Worst-Case Expert Migration Message Size and Latency Per GPU ($48 \times E \times d_{\text{model}} \times d_{\text{ffn}} / G$ bytes, $G=8$, bandwidth $=50$ GB/s)}
\label{tab:migration_worst_case}
\begin{tabular}{lrrrrr}
\toprule
\textbf{Model} & \textbf{E/layer} & \textbf{$d_{\text{model}}$} & \textbf{$d_{\text{ffn}}$} & \makecell{\textbf{Send Size /}\\\textbf{GPU (GB)}} & \makecell{\textbf{Latency}\\\textbf{(ms)}} \\
\midrule
Switch-Base          & 128 & 768   &  2{,}048 &   1.21 &   24.2 \\
Mixtral 8$\times$7B  &   8 & 4{,}096 & 14{,}336 &   2.63 &   52.6 \\
Mixtral 8$\times$22B &   8 & 6{,}144 & 16{,}384 &   4.50 &   90.0 \\
Grok-1               &   8 & 6{,}144 & 32{,}768 &   9.00 &  180.0 \\
GLaM (1.2T)          &  64 & 8{,}192 & 32{,}768 & 102.88 & 2057.6 \\
DeepSeek-V2          & 160 & 5{,}120 &  1{,}536 &   7.04 &  140.8 \\
DeepSeek-V3          & 256 & 7{,}168 &  2{,}048 &  21.00 &  420.0 \\
\bottomrule
\end{tabular}
\label{tab:migration-cost}
\end{table}

With average intra-node all-to-all bandwidth 50GB/s, the worst case expert migration will only take tens of milliseconds for most models. However, we will perform this migration incrementally and intermittently so that only a subset of the experts need to migrate to balance GPU workload. To rebalance with minimal swaps, we develop a hill-climbing swapping based algorithm (Algorithm~\ref{alg:rebalance}).

\begin{algorithm}[!htbp]
\caption{Hill-Climbing Swap-Based Minimal Rebalancing}
\label{alg:minimal_rebalance}
\KwIn{Groups $\mathcal{G} = \{G_1, \ldots, G_K\}$, max iterations $T = 100$}
\KwOut{Rebalanced groups $\mathcal{G}$, swap count $c$}

$c \leftarrow 0$\;

\For{$t = 1, \ldots, T$}{
    $s_k \leftarrow \sum_{n \in G_k} n$ \quad $\forall k$\;
    $k^+ \leftarrow \arg\max_k\, s_k$\;
    $k^- \leftarrow \arg\min_k\, s_k$\;
    $\delta \leftarrow s_{k^+} - s_{k^-}$\;
    $\text{best\_swap} \leftarrow \texttt{None}$, \quad $\Delta^* \leftarrow 0$\;

    \ForEach{$(i, n_1)$ in $G_{k^+}$}{
        \ForEach{$(j, n_2)$ in $G_{k^-}$}{
            $\delta' \leftarrow |(s_{k^+} - n_1 + n_2) - (s_{k^-} - n_2 + n_1)|$\;
            \If{$\delta' < \delta$ \textbf{and} $(\delta - \delta') > \Delta^*$}{
                $\Delta^* \leftarrow \delta - \delta'$, \quad $\text{best\_swap} \leftarrow (i, j)$\;
            }
        }
    }

    \eIf{$\text{best\_swap} \neq \texttt{None}$}{
        $(i^*, j^*) \leftarrow \text{best\_swap}$\;
        $G_{k^+}[i^*] \leftrightarrow G_{k^-}[j^*]$\;
        $c \leftarrow c + 1$\;
    }{
        \textbf{break}\;
    }
}
\Return{$\mathcal{G},\, c$}
\label{alg:rebalance}
\end{algorithm}

\section{Scalable Training of SOTA MoE Models}\label{sec:executor}
\begin{figure}[!htbp]
    \centering
    \includegraphics[width=0.95\linewidth]{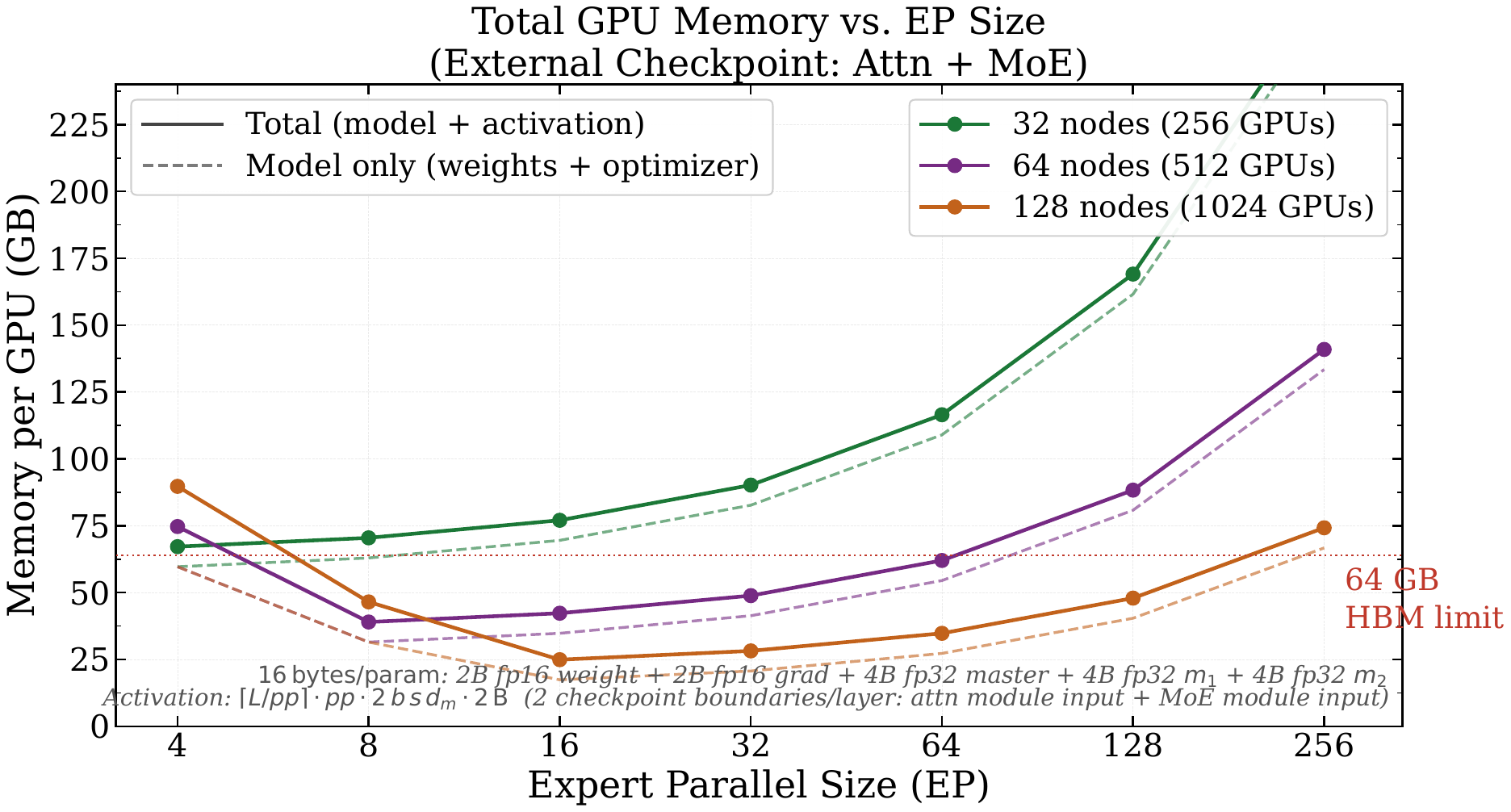}
    \caption{Identifying viable training strategies for a model by estimating its per-GPU memory requirement. Here, we investigate training strategy of a 615 B parameter model. And, just like this analysis suggests, we can train this model with at least 64 nodes.}
    \label{fig:training-strategy}
\end{figure}

We have developed resource modeling and a throughput estimator through micro-benchmarking of Frontier supercomputer to identify the viable and performant distributed training strategies. For example to train a model (super model with ~545B parameters), we computed the required memory for various node count with different distribution strategies (Figure~\ref{fig:training-strategy}).

\subsection{Trroughput Ceiling : Training a Single Layer in a Node}

Every transformer layer has two major components, non-expert attention module, and multiple Feed Forward Network (FFN) experts. In an expert-data parallel setting, we need to distribute the experts equally across the G gpus in a node and replicate the attention module in a data-parallel way across all G gpus. Each GPU needs to host and process one attention module and $\dfrac{E}{G}$ experts. 

\begin{figure}[!htbp]
    \centering
    \includegraphics[width=0.8\linewidth]{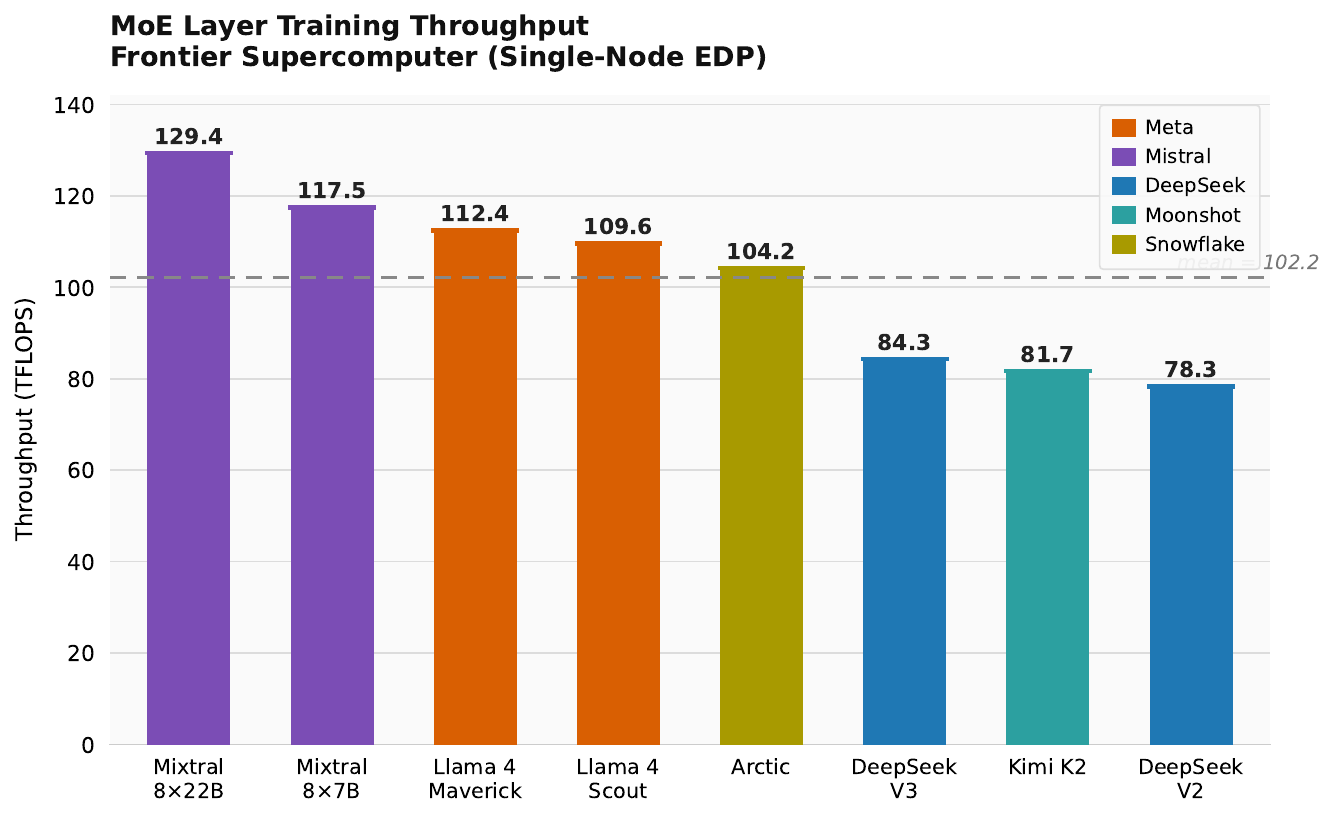}
    \caption{Training efficiency of a single layer of the SOTA models.}
    \label{fig:sota-single-layer}
\end{figure}

After getting the suitable configuration, we train a single layer of the SOTA models on a single Frontier node using expert-parallelism and observe their training performance (Figure~\ref{fig:sota-single-layer}). We were able to fit a single layer to a single node for each of the models we tried. With memory constraint under pipelined expert-parallelism
and pipeline bubble under various schedules, this sets up the upper limit of training performance of any model using Piper.

\subsection{Training the Full Model}
\begin{figure}[t]
    \centering
    \includegraphics[width=1.0\columnwidth]{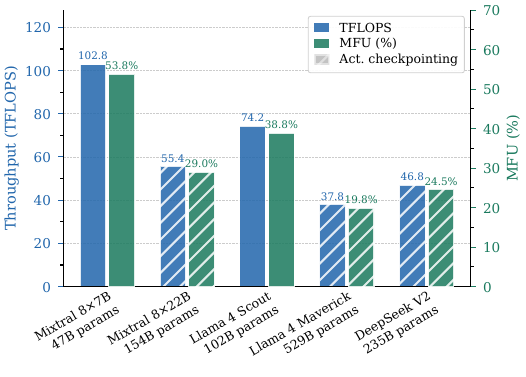}
    \caption{Training throughput of SOTA MoE models using Piper
             (sequence length 4096). Hatching denotes activation
             checkpointing.}
    \label{fig:sota-training}
\end{figure}

With our resource modeling, we identified valid parallelization strategies for various SOTA models and train them using their original size. We selectively use activation checkpointing to fit the model into a small number of GPUs while minimizing the impact on the MFU (Figure~\ref{fig:sota-training}). From Table~\ref{fig:sota-training}, we observe that training models with traditional large experts achieves best performance while the fine-grained experts achieve lower performance.

\subsection{Comparison Against Other MoE Training Frameworks}
\begin{figure}[!htbp]
    \centering
    \includegraphics[width=1.0\linewidth]{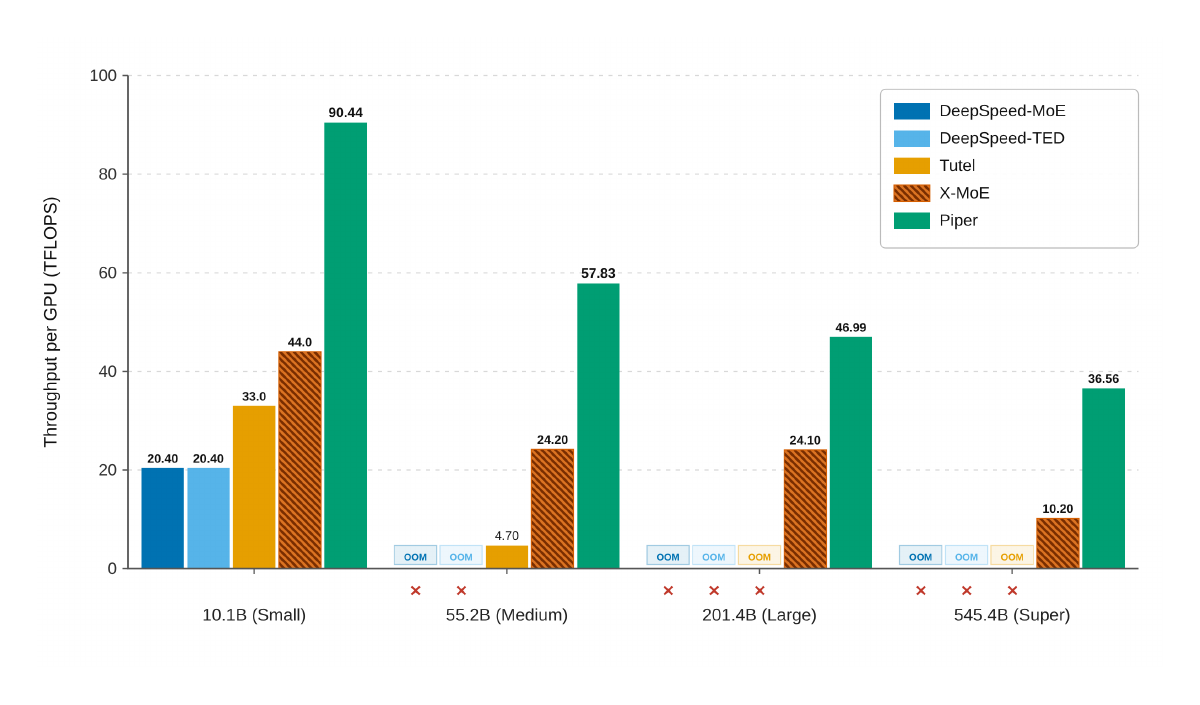}
    \caption{Throughput comparison against SOTA MoE training frameworks. Piper can train the small, medium, large, and super model using 8, 32, 80, and 512 MI250X GPUS instead of 256 and 1024 GPUs used by X-MoE.}
    \label{fig:framework-comparison}
\end{figure}

We compare Piper's training throughput for different model sizes against state of the art training frameworks such as Tutel, DeepSpeed MoE, DeepSpeed TED, and most prominent X-MoE. Since X-MoE is the leading open-source framework for fine-grained experts training, we compare Piper against these tools using fine-grained MoEs. Our framework Piper can train these models with only a fraction of their requirement at 2-3.6X throughput (Figure~\ref{fig:framework-comparison}).

\subsection{Training Trillion(s) Parameter Models}
\begin{figure}[!htbp]
    \centering
    \includegraphics[width=0.9\columnwidth]{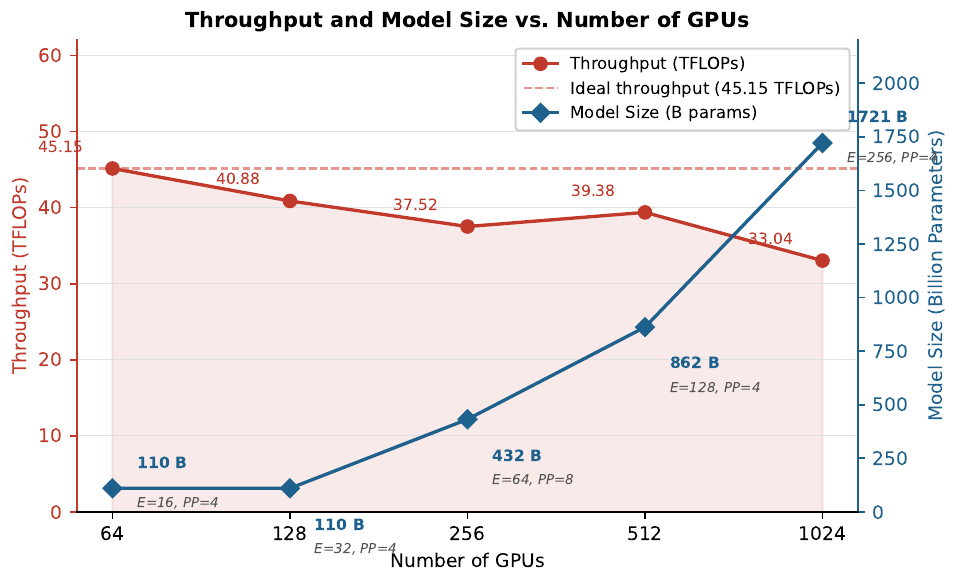}
    \caption{Scaling M10B models with experts.}
    \label{fig:scaling-with-experts}
\end{figure}

We take a dense model $[d_{model} = 5120, d_{ffn} = 20480, L = 32, k=2]$ (10 Billion parameters, let's call it M10B) and scale out its parameter count by scaling the number of experts ($E$). We start with 16 experts on 8 nodes and then scale $E$ proportionally. We use 128 experts on 64 nodes (512 GPUs), and 256 experts on 128 nodes (1024 GPUs). With this scaling, we can train a 862 Billion parameter model using 512 GPUs at 39.38 TFLOPs, and a 1.7 Trillion parameter model using 1024 GPUs at 33 TFLOPS. This is a form of weak scaling, and the scaling efficiency is 73\% from 64 GPUs to 1024 GPUs (Figure~\ref{fig:scaling-with-experts}).

\section{Conclusion and Discussion}
We presented Piper, a holistic MoE training framework that co-designs distributed training strategy through mathematical resource modeling, empirical micro-benchmarking, and platform-aware performance estimation. Its central contribution is the application of pipeline parallelism on top of traditional expert-data parallelism, which proves critical in mitigating large-scale all-to-all latency. Piper achieves 2–3.6× the MFU of X-MoE on SOTA models, demonstrating that deep understanding of the target HPC platform translates directly into training throughput gains. Implemented in Python and PyTorch, Piper is portable across platforms, and its micro-benchmarking suite can characterize arbitrary HPC systems. As new MoE architectures emerge, Piper's modular design allows incremental integration of kernel optimizations, improved schedules, and dynamic load-balancing strategies to deliver continued efficiency gains.

\clearpage
\bibliographystyle{IEEEtran}
\bibliography{references}

\end{document}